\documentclass[preprint]{aastex}
\usepackage{natbib}
\usepackage{booktabs}
\usepackage{threeparttable}

\newcommand{\siiv}{Si \scriptsize{IV} \normalsize}

\begin{document}

\title{Quasi-periodic pulsation detected in Lyman-alpha emission during solar flares}

\author{Dong~Li\altaffilmark{1,2,3}, Lei~Lu\altaffilmark{1}, Zongjun~Ning\altaffilmark{1}, Li~Feng\altaffilmark{1}, Weiqun~Gan\altaffilmark{1}, and Hui~Li\altaffilmark{1} }
\affil{$^1$Key Laboratory of Dark Matter and Space Astronomy, Purple Mountain Observatory, CAS, Nanjing 210033, People¡¯s Republic of China \\
     $^2$State Key Laboratory of Space Weather, Chinese Academy of Sciences, Beijing 100190, People¡¯s Republic of China \\
     $^3$CAS Key Laboratory of Solar Activity, National Astronomical Observatories, Beijing 100101, People¡¯s Republic of China  \\}
     \altaffiltext{}{Correspondence should be sent to: lidong@pmo.ac.cn \& leilu@pmo.ac.cn}

\begin{abstract}
We investigated the quasi-periodic pulsation (QPP) in Ly-$\alpha$,
X-ray and extreme-ultraviolet (EUV) emissions during two solar
flares, i.e., an X-class (SOL2012-01-27T) and a C-class
(SOL2016-02-08T). The full-disk Ly-$\alpha$ and X-Ray flux during
these solar flares were recorded by the EUV Sensor and X-Ray Sensor
on board the {\it Geostationary Operational Environmental
Satellite}. The flare regions were located from the EUV images
measured by the Atmospheric Imaging Assembly. The QPP could be
identified as a series of regular and periodic peaks in the light
curves, and its quasi-periodicity was determined from the global
wavelet and Fourier power spectra. A quasi-periodicity at about
3~minutes is detected during the impulsive phase of the X-class
flare, which could be explained as the acoustic wave in the
chromosphere \cite[e.g.,][]{Milligan17}. Interestingly, a
quasi-periodicity at roughly 1~minute is discovered during the
entire evolutionary phases of solar flares, including the precursor,
impulsive, and gradual phases. This is the first report of 1-minute
QPP in the Ly-$\alpha$ emission during solar flares, in particular
during the flare precursor. It may be interpreted as a
self-oscillatory regime of the magnetic reconnection, such as
magnetic dripping.

\end{abstract}

\section{Introduction}
Quasi-periodic pulsation (QPP) is a frequently observed feature
during flare emissions on the Sun or Sun-like stars. A typical QPP
is often regarded as the temporal and regular fluctuation of
electromagnetic radiation in solar/stellar flares \citep[see,][for
reviews]{Nakariakov09,Van16,McLaughlin18}, which conveys the
temporal feature and plasma characteristics of the flare radiation.
Therefore, it should be useful to diagnose the coronal parameters of
the Sun or remote Sun-like stars \citep[e.g.,][]{Pugh19,Yuan19}. It
was first found in solar X-ray emission \citep{Parks69}, then more
and more QPP events are discovered in solar/stellar radiation
intensity or flux over a broad range of wavelengths, i.e., radio
emission
\citep[e.g.,][]{Ning05,Kolotkov15,Kupriyanova16,Nakariakov18},
H$\alpha$ emission \citep[e.g.,][]{Srivastava08,Yang16}, Ly-$\alpha$
emission \citep[e.g.,][]{Milligan17,Milligan19}, extreme-ultraviolet
(EUV) wave bands \citep[e.g.,][]{Yuan11,Yuan19,Li18,Hayes19}, and
soft or hard X-ray (SXR or HXR) channels
\citep[e.g.,][]{Asai01,Foullon05,Ofman06,Li08,Ning14,Hayes16}.
Moreover, the QPP has also been measured in spectroscopic
observations, i.e. the regular and repeating variations in the
Doppler velocity, line width and intensity of spectral lines
\citep[e.g.,][]{Li15,Wang15,Tian16,Brosius18}. Finally, the QPPs in
solar flares are found to share some main frequencies with the
quasi-periodic fast-propagating (QFP) magnetosonic waves
\citep[e.g.,][]{Liu11,Shen12a,Nistic14,Shen18a}, indicating their
common origin \citep{Yuan13,Liu14,Shen18b,Kumar17}.

A typical QPP event can oscillate a few to tens of cycles, that is,
the QPP can decay quickly \citep{Wang03,Anfinogentov13} or could be
persistent \citep[decay-less,][]{Tian12,Tian16}. On the other hand,
the oscillation period can be observed from milliseconds through
seconds to dozens of minutes
\citep{Aschwanden94,Schrijver02,Tan10,Inglis16,Lid17,Kolotkov18,Shen18c}.
Sometimes, the QPP with a similar period has been detected in a wide
range of wave bands \citep{Dolla12,Lid15}, while the QPP with
multi-periods has also been found in a same flare event
\citep{Inglis09,Zimovets10,Chowdhury15,Kolotkov15}. More
specifically, the period ratio typically deviates from two, which
might be attributed to the highly dispersive magnetohydrodynamic
(MHD) mode, or the stratification of plasma density
\citep[e.g.,][]{Andries05}, or the expansion of flaring loop
\citep[e.g.,][]{Verth08}. The signature of QPP can be discovered in
all phases of solar flares, such as precursor \citep{Tan16},
impulsive phase \citep{Li17} and gradual phase \citep{Lid18b}.
Moreover, different periods have been already found to be presented
at different phases of the same flare event
\citep{Hayes16,Dennis17}.

Statistical study shows that 80\% of the X-class flares exhibit the
signature of QPPs in {\it GOES} X-ray channels during the impulsive
phase \citep{Simoes15}. However, the generation mechanism of the QPP
is still highly debated, which strongly depends on the observed
period and wavelength
\citep{Aschwanden94,Nakariakov09,Inglis12,McLaughlin18}. Generally,
the QPP detected in nonthermal emission such as microwave or HXR
channel is thought to be related to the electron beam, which is
accelerated by a periodic process of energy release during flare
eruptions \citep{Kliem00,Asai01,Inglis09,Murray09}. On the other
hand, the QPP with a short period observed in radio emission is
often associated with the dynamic interaction between energetic
particles and waves, while a long-period QPP seen in the white light
or EUV wave bands is supposed to be associated with the dynamic of
emitting plasmas in solar active region or the whole Sun
\citep{Aschwanden87,Chen06,Nakariakov06,Nakariakov09}. Briefly, the
QPP can be directly driven by a magnetohydrodynamic (MHD) wave in
slow, kink, or sausage modes
\citep{Nakariakov06,Nakariakov16,Reznikova11,Van16}, it could also
be caused by a process of repetitive magnetic reconnection which
might be either spontaneous (such as magnetic dripping) or induced,
i.e., by MHD oscillations
\citep{Murray09,McLaughlin09,McLaughlin18,Thurgood17}.

Lyman-alpha (Ly-$\alpha$) is a neutral hydrogen line at 1216~{\AA}
in the chromosphere, and it is the most prominent emission line in
solar UV spectrum during flare eruptions
\citep{Woods04,Allred05,Milligan12,Milligan14,Milligan16}. It is
believed that the fluctuations in Ly-$\alpha$ emission can cause the
changes in planetary atmospheres during the high-activity period.
Therefore, it is important to investigate the variability of
Ly-$\alpha$ emission during solar flares, which may be helpful for
assessing its relative effect
\citep[see,][]{Milligan12,Milligan19,Milligan16}. However, it is not
extensively studied of the temporal variation in Ly-$\alpha$
emission during solar flares since it was first reported by
\cite{Canfield80}, which is due to the limitation of the
observational instruments. There are apparently very few studies of
QPPs related to the Ly-$\alpha$ emission. \cite{Milligan17} first
detected the QPP with a period of about 3~minutes in full-disk
Ly-$\alpha$ and Lyman continuum (LyC) emission during the impulsive
phase of an X2.2 flare. Then they reported a 4.4-minute QPP in
full-disk Ly-$\alpha$ emission during the impulsive phase of an X1.1
flare. Both of these two QPPs are attributed to the acoustic waves
induced by solar flares \citep{Milligan19}.

The one-minute QPP has been detected in full-disk SXR flux during a
solar flare \citep[e.g.,][]{Ning17,Hayes19}, and the QPP with
periods of 3 and 4.4 minutes has been detected in the full-disk
Ly-$\alpha$ emission \citep{Milligan17,Milligan19} during two
flares. In this paper, we discovered the 1-minute QPP in
Ly-$\alpha$, EUV, and X-ray emission during two solar flares. Our
finding could provide an observational constraint for the QPP
generation mechanism. The observations in this paper are taken from
the joint instruments, i.e., the EUV Sensor
\citep[EUVS,][]{Viereck07} and the X-Ray Sensor
\citep[XRS,][]{Hanser96} aboard the {\it Geostationary Operational
Environmental Satellite} ({\it GOES}), and the Atmospheric Imaging
Assembly \citep[AIA,][]{Lemen12} on board the {\it Solar Dynamics
Observatory} ({\it SDO)}.

\section{Observations}
Two solar flares are selected to investigate the QPP in this paper,
both of which show a prominent QPP feature in the full-disk
Ly-$\alpha$ emission. One is an X1.7 flare on 2012 January 27, and
takes place on the active region of $NOAA$~11402. The other is a
C1.6 flare on 2016 February 8, occurring in the active region of
$NOAA$~12492. Both of them are well recorded by the {\it GOES}/XRS,
{\it GOES}/EUVS, and {\it SDO}/AIA, as seen in Table~\ref{tabl}.

Figure~\ref{flare1}~(a)$-$(d) shows the EUV images in AIA~94~{\AA}
and 304~{\AA} at around 17:45~UT and 18:45~UT on 2012 January 27,
respectively. The flare occurs on the solar limb. The magenta box
outlines the local region which is used to integrate the flare flux
in AIA~94~{\AA} and 304~{\AA}. Due to the saturation effect during
the flare period, only short-exposure images with a temporal cadence
of $\sim$24~s are used here \citep{Lemen12,Ning17}. Panel~(e)
presents the full-disk light curves in Ly-$\alpha$ (black),
SXR~1-8~{\AA} (red) and 0.5$-$4~{\AA} (blue) emission recorded by
the {\it GOES}. The {\it GOES} 1$-$8~{\AA} flux shows that an X1.7
flare starts to increase at roughly 18:05~UT, which could be
identified as the onset time of this flare. On the other hand, a
small but well-developed SXR peak appears at around 17:50~UT (red
arrow), which is a bit earlier than the onset time of solar flare.
The small pulse can also be found in {\it GOES}~0.5$-$4~{\AA} and
Ly-$\alpha$ line flux, which is more prominent. However, these three
{\it GOES} light curves are measured from the whole Sun. To
determine if the small pulse is related to the X1.7 flare, we then
plot the local EUV flux using the spatially resolved {\it SDO}/AIA
observations. Similar to the full-disk {\it GOES} flux, the local
light curves in AIA~94~{\AA} (green) and 304~{\AA} (orange) also
shows the small pulse before the flare onset time. In particular,
the pulse in Ly-$\alpha$ emission and AIA~304~{\AA} is a little
earlier but much more pronounced than that in {\it GOES} SXR
channels and AIA~94~{\AA}. Therefore, the small pulse occurring just
before the flare onset could be considered as a flare precursor
\citep[][]{Benz17,Lid18c,Battaglia19,Shen19}, which can be caused by
an external reconnection as described in the magnetic breakout model
\citep[see][]{Shen12b,Chen16}. Moreover, it is thought to play an
important role in triggering the major flare
\citep[e.g.][]{Priest02,Lin05,Zhou19}.

Figure~\ref{flare2}~(a)$-$(d) presents AIA EUV images in 94~{\AA}
and 304~{\AA} at about 05:10~UT and 05:30~UT on 2016 February 8. The
flare takes place near the solar disk center. Two magenta boxes (L
\& S) mark two bright regions used to integrate the local EUV
fluxes. The flare is consisted of a circular flare ribbon and a
bright inner flare ribbon, which is associated with negative
magnetic polarities (blue contours) outside and positive magnetic
polarities (yellow contours) inside, as shown in the solid box (L)
in panel~(d). Here the positive and negative magnetic fields were
derived from the line-of-sight magnetogram that measured by the
Helioseismic and Magnetic Imager \citep{Schou12}. Thus, the C1.6
flare can be identified as a circular-ribbon flare
\citep{Ichimoto84,Masson09,Zhang16}. Panel~(e) shows the full-disk
{\it GOES} flux in Ly-$\alpha$ (black) and SXR (red \& blue)
emission. Similar to the X1.7 flare, the {\it GOES} light curves
also show a small and well-developed peak before the flare onset
time such as $\sim$05:22~UT. Then we plot the local EUV light curves
integrated over a large bright region (L) in AIA~94~{\AA} (solid
green) and 304~{\AA} (orange). They both exhibit the flare peak
after $\sim$05:22~UT, but they lack the small peak before the flare
onset time. On the other hand, the small peak can be found in a
small bright region (S) in AIA~94~{\AA}, as shown by a green dashed
line. The observational fact suggests that the small pulse in SXR
channels could be a solar microflare \citep{Hannah11,Nakariakov18}
rather than a flare precursor.

\section{Data reduction and Results}
To examine the QPP feature in these two flares, the fast Fourier
transform method \citep[see,][]{Ning17} is performed to the original
light curves to obtain the Fourier spectrum in {\it GOES}
Ly-$\alpha$ and SXR channels, as shown in the panels (f) \& (g) of
Figures~\ref{flare1} and \ref{flare2}. In the astrophysical
observations \citep{Vaughan05,Pugh17,Wang20}, the term `red noise'
is often described by a power-law model in the Fourier power
spectrum at longer periods, i.e., $P(T)~\sim~T^{\theta}$, where $T$
is the period, $\theta$ represents a slope in the log-log coordinate
system. While a white noise often refers to the flat spectrum at the
shorter-period end. In the solar atmosphere, it is a very common
observational phenomenon of such superposition of red and white
noises, which are dominated at longer and shorter periods,
respectively
\citep[e.g.,][]{Inglis15,Kolotkov16,Ning17,Li20,Liang20}.

Figure~\ref{flare1}~(f) and (g) presents Fourier power spectra of
the X1.7 flare in Ly-$\alpha$ and {\it GOES}~1$-$8~{\AA} channels.
The double peaks at roughly 1 and 3 minutes (cyan and pink arrows)
are found to exceed the confidence level of 99\% (magenta line) in
these two power-law spectra, suggesting two periods in this
flare-related QPP. We also notice much longer periods
($>$10~minutes) appearing in {\it GOES}~1$-$8~{\AA}, which is out of
the scope of this study, because it is absent in Ly-$\alpha$
emission. Figure~\ref{flare2}~(f) and (g) shows a similar Fourier
power spectrum but for the C1.6 flare, and only one peak at around 1
minute can be detected to be above the 99\% confidence level in the
two power-law spectra, implying a 1-minute QPP in this C1.6 flare.
The 3-minute QPP in Ly-$\alpha$ emission has been reported by
\cite{Milligan17}. Therefore, we focus on the QPP with a shorter
period of $\sim$1~minute in this paper.

To investigate the 1-minute QPP of these two flares in detail, the
wavelet analysis \citep[e.g.,][]{Torrence98} is performed on the
detrended light curve after removing a $\sim$110 s running average
\citep[see,][]{Yuan11,Tian16}. The detrended light curve is used
here because we thereby enhance the short-period QPP such as $\sim$1
minute and suppress the long-period trend, the discussion and
justification of this method has been reported in detail by
\cite{Gruber11,Kupriyanova10,Auchere16,Dominique18} etc.

Figure~\ref{lya1} shows the wavelet analysis result of the X1.7
flare in the full-disk Ly-$\alpha$ emission. Panel~(a) displays the
normalized light curve (black) in the Ly-$\alpha$ emission from
$\sim$17:30~UT to $\sim$20:00~UT on 2012 January 27, and its trended
light curve is overplotted with a magenta line. Panel~(b) gives the
normalized detrended light curve, and it is characterized by a
series of repeat and regular peaks, which could be regarded as the
signature of QPP. Then, the oscillation period can be determined
from the wavelet power spectrum and the global wavelet power, as
shown in panels (c) and (d). They both exhibit double periods, i.e.,
a shorter period of $\sim$1~minute and a longer period of
$\sim$3~minutes, which are consistent with the Fourier analysis
result in Figure~\ref{flare1}~(f). Moreover, the 1-minute QPP can be
discovered between $\sim$17:35~UT$-$19:40~UT, which is ranging from
the flare precursor through the impulsive phase to the gradual
phase. However, the 3-minute QPP only appears during the impulsive
phase ($\sim$18:05~UT$-$18:30~UT), which is similar to previous
findings \citep[e.g.,][]{Milligan17,Milligan19}.

Figure~\ref{goes1} presents the wavelet analysis result in the
full-disk SXR~1$-$8~{\AA} flux of the X1.7 flare. The left panels
show the normalized light curve (a), the normalized detrended light
curve (c) and the wavelet power spectrum (e) from $\sim$17:30~UT to
$\sim$20:00~UT. It can be seen that both the 1- and 3-minute QPPs
tend to appear between $\sim$18:10$-$18:50. There is not any
apparent signature of QPP before 18:00~UT, which could be attributed
to the fact that the SXR radiation during the solar flare is much
stronger than that during the flare precursor, as shown in
panel~(a). A series of repeat but very small peaks can be found
between $\sim$17:35$-$17:50, as indicated with a cyan arrow in
panel~(c). Therefore, the SXR flux during the flare precursor is
selected to perform the wavelet analysis, as shown in the right
panels. The original and detrended light curves are given in
panels~(b) and (d), and the QPP feature is pronounced in the
detrended light curve. The 1-minute QPP can be found between
$\sim$17:35~UT and $\sim$17:50~UT in the wavelet power spectrum, as
shown in panel~(f).

\begin{figure}
\centering
\includegraphics[width=\linewidth,clip=]{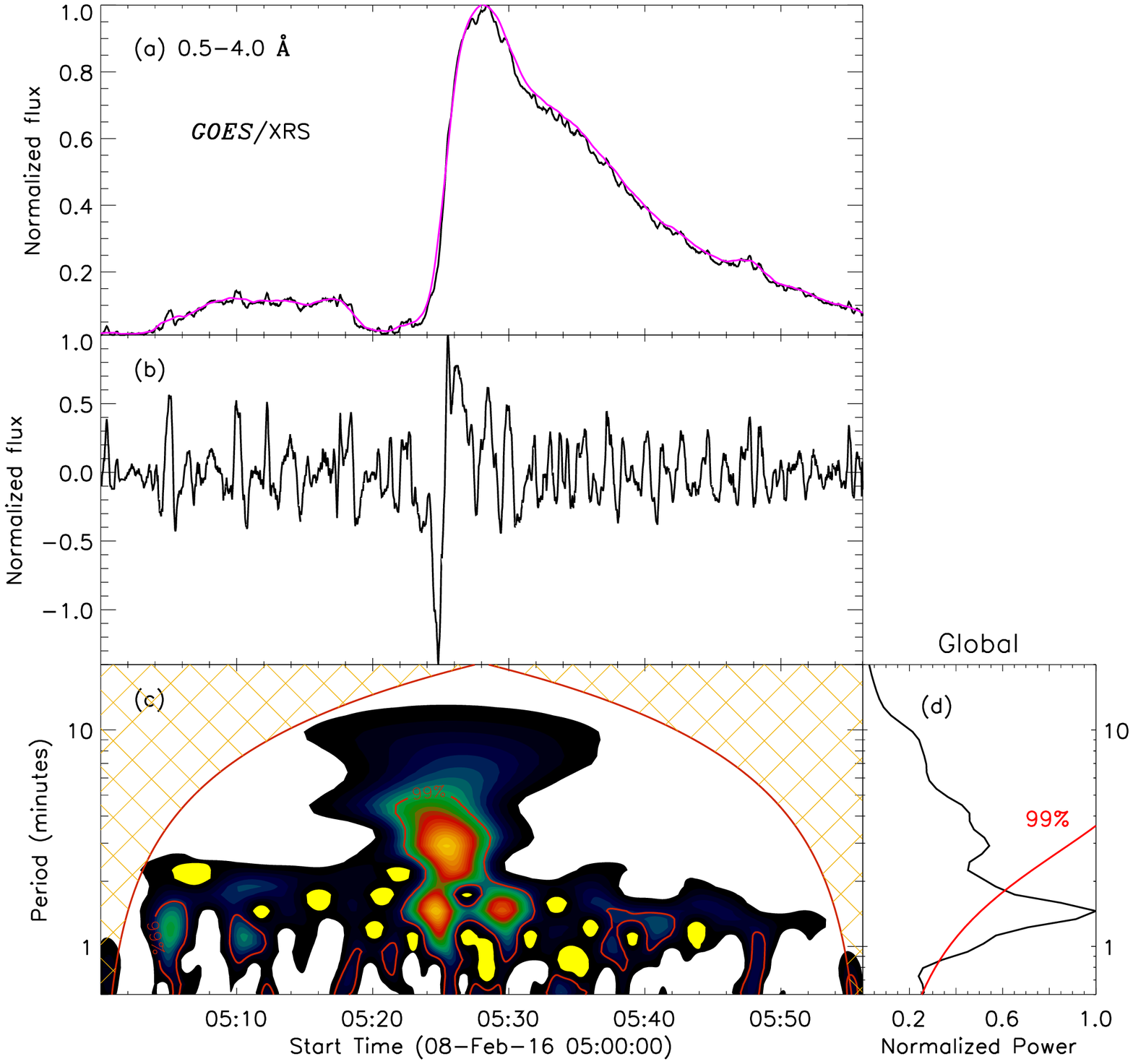}
\caption{Wavelet analysis result in {\it GOES}~0.5$-$4~{\AA} of the
C1.6 flare. Panel~(a): Normalized original (black) and trended
(magenta) light curves. Panel~(b): Normalized detrended light curve.
Panels~(c)~\&~(d): The wavelet power spectrum and the global wavelet
power. The red lines indicate significance levels of 99\%.
\label{goes2}}
\end{figure}

Figure~\ref{lya2} shows the similar wavelet analysis result of the
C1.6 flare in the full-disk Ly-$\alpha$ emission. The normalized
original (black) and trended (magenta) light curves from 05:00~UT to
05:56~UT are given in panel~(a). Panel~(b) presents the normalized
detrended light curve, which exhibits a pronounced QPP feature. The
1-minute period can appear simultaneously in the wavelet power
spectrum (c) and the global wavelet power (d), and it could remain
from the impulsive to gradual phases, i.e., from $\sim$05:15~UT to
$\sim$05:45~UT. We also notice that the 1-minute period can be found
between around 05:05~UT$-$05:10~UT, which is believed as a
microflare. However, the wavelet power, as well as the Ly-a
emission, is relatively weak. Thus, it is not considered in this
work. The similar wavelet analysis result can be derived from the
full-disk SXR flux measured by {\it GOES}/XRS, as shown in
Figure~\ref{goes2}. It exhibits a strong QPP with a period of
roughly 1~minute during the impulsive phase, i.e., between
$\sim$05:20~UT$-$05:30~UT. It also displays a weak signature of
1-minute QPP during the gradual phase. We also notice that the onset
time of QPP in SXR channel is later than that seen in Ly-$\alpha$
emission, which is attributed to the time delay between the SXR flux
and the Ly-$\alpha$ irradiance \citep[see
also][]{Milligan16,Chamberlin18}.

The 1-minute QPP is discovered in the full-disk light curves in the
Ly-$\alpha$ and SXR channels. Thanks to the high-spatial resolution
imaging observations from {\it SDO}/AIA, we can obtain the local EUV
flux such as 304~{\AA}. Figure~\ref{aia} shows the Fourier spectra
in local AIA~304~{\AA} of the X1.7 (a) and C1.6 (b) flares,
respectively. Both of these two flares display a pronounced QPP
feature at a period of roughly 1~minute, but only the X1.7 flare
appear the 3-minute QPP. All these observational results based on
the local EUV flux agree well with previous findings derived from
the full-disk light curves recorded by the {\it GOES}. Finally, the
ARIMA model and the SARIMA model with a periodic component
\citep[e.g.,][]{Box15,Hyndman18} are applied to the light curve of
the solar flare, respectively. The Akaike information criterion
(AIC) of the latter was better than the former model, suggesting
that the true periodicity is present in this work.

\section{Conclusion and Discussion}
Using the {\it GOES}/EUVS observations, we discovered the 1-minute
QPP in the Ly-$\alpha$ emission during two solar flares, i.e., an
X1.7 flare, and a C1.6 circular-ribbon flare. Moreover, the 1-minute
QPP can be detected during the whole flare phases, such as flare
precursor, impulsive and gradual phases. On the other hand, we could
not detect the similar 1-minute QPP in Ly-$\alpha$ light curves
before and after the two flares under study, which indicates that
the 1-minute QPP mainly comes from the flare radiation rather than
the background emission of the Sun. It is well known that the
neutral hydrogen Ly-$\alpha$ is the strongest emission line in the
solar UV spectrum, which shows a significant enhancement during
solar flares \citep[e.g.,][]{Woods04,Milligan12,Milligan19}.
However, the Ly-$\alpha$ variability of solar flares has not yet
been investigated extensively, mainly due to the limited
observational instruments \citep{Kretzschmar13,Milligan19}.
Therefore, the QPP in Ly-$\alpha$ emission is rarely studied.
\cite{Milligan17} first found the 3-minute QPP in full-disk
Ly-$\alpha$ and LyC emission during the impulsive phase of an X2.2
flare, which was attributed to be the acoustic waves in the
chromomere. Later on, they discovered another 4.4-minute QPP in
Ly-$\alpha$ emission from the full-disk irradiance during the
impulsive phase of an X1.1 flare at solar limb \citep{Milligan19}.
In our studies, a similar 3-minute QPP is observed during the
impulsive phase of the X1.7 flare. However, it failed to be detected
in the C1.6 flare, which might be due to its weak radiation and
short lifetime.

The QPP with a period of around 1 minute can also be found in X-ray
and EUV wave bands, such as {\it GOES}~1$-$8~{\AA}, 0.5$-$4~{\AA},
and AIA~304~{\AA}. The 1-minute QPP has been extensively studied
during an X1.6 flare on 2014 September 10 in multiple wavelengths,
such as SXR, HXR and EUV wave bands, and it can be found during the
impulsive and gradual phases \citep{Ning17}. On the other hand, the
1-minute QPP was found in SXR flux during the impulsive phase of an
X8.2 flare on 2017 September 10 \citep{Hayes19}. Besides that, we
further detected the 1-minute QPP in Ly-$\alpha$ emission during the
precursor of an X-class flare. The QPP with a period of 32$-$42~s
has been reported during a C3.1 circular-ribbon flare in the
\siiv~line intensity and SXR derivative flux \citep{Zhang16}. In
this paper, we found the 1-minute QPP in Ly-$\alpha$ line and
SXR/EUV flux during a C-class circular-ribbon flare.

\begin{figure}
\centering
\includegraphics[width=\linewidth,clip=]{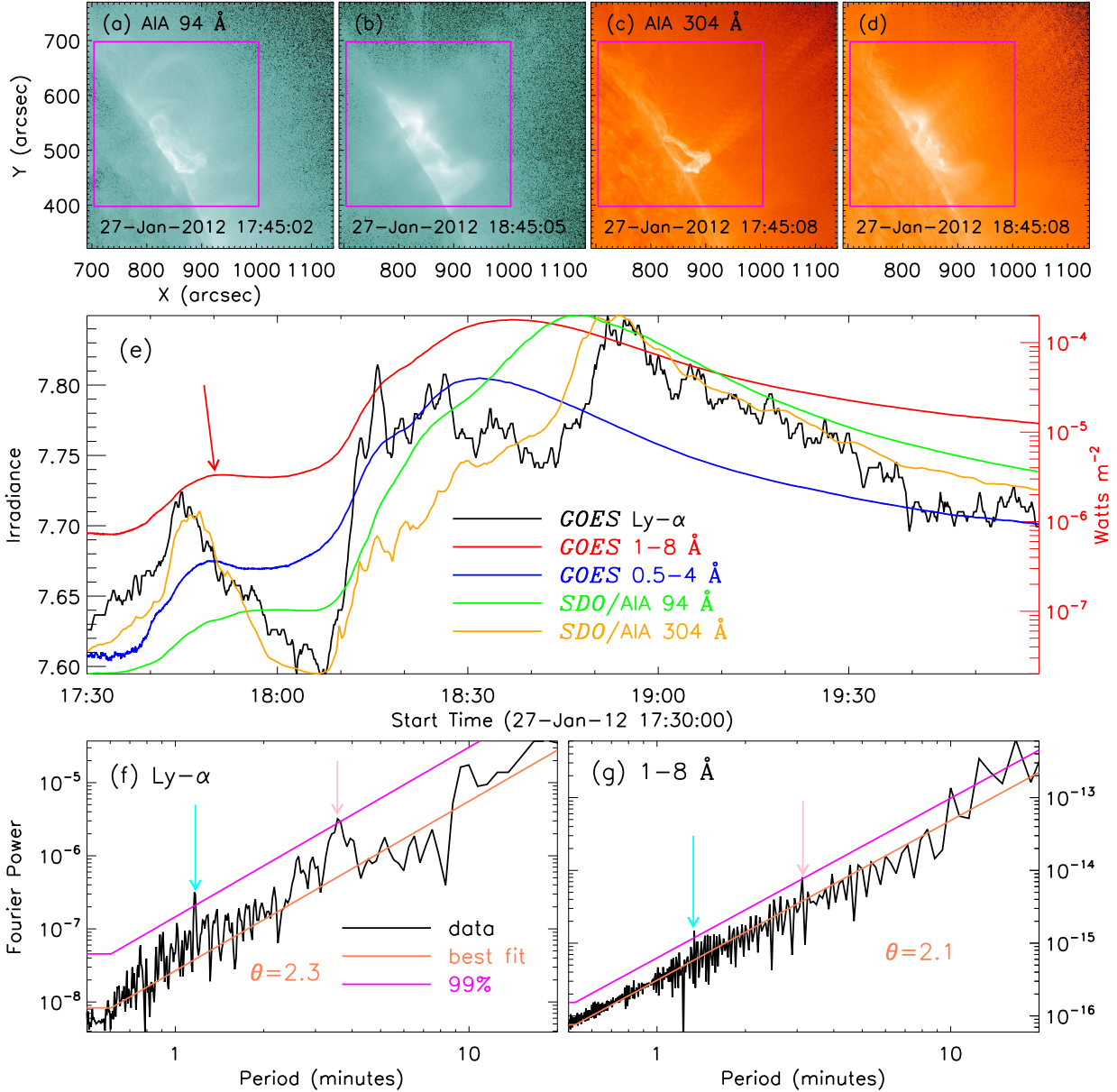}
\caption{Overview of the solar flare on 2012 January 27.
Panels~(a)$-$(d): AIA images in 94~{\AA} and 304~{\AA}. The magenta
box marks the flare region used to integrate the local EUV flux.
Panel~(e): Full-disk flux in Ly-$\alpha$ (black), {\it
GOES}~1$-$8~{\AA} (red), and 0.5$-$4.0~{\AA} (blue), the local flux
in AIA~94~{\AA} (green) and 304~{\AA} (orange). Panels~(f)$-$(g):
Fourier power spectra derived from the Ly-$\alpha$ and {\it
GOES}~1$-$8~{\AA}. The coral and magenta lines indicate the best fit
results and the confidence levels at 99\%, respectively. The cyan
and pink arrows indicate the periods above the confidence level.
\label{flare1}}
\end{figure}

\begin{figure}
\centering
\includegraphics[width=\linewidth,clip=]{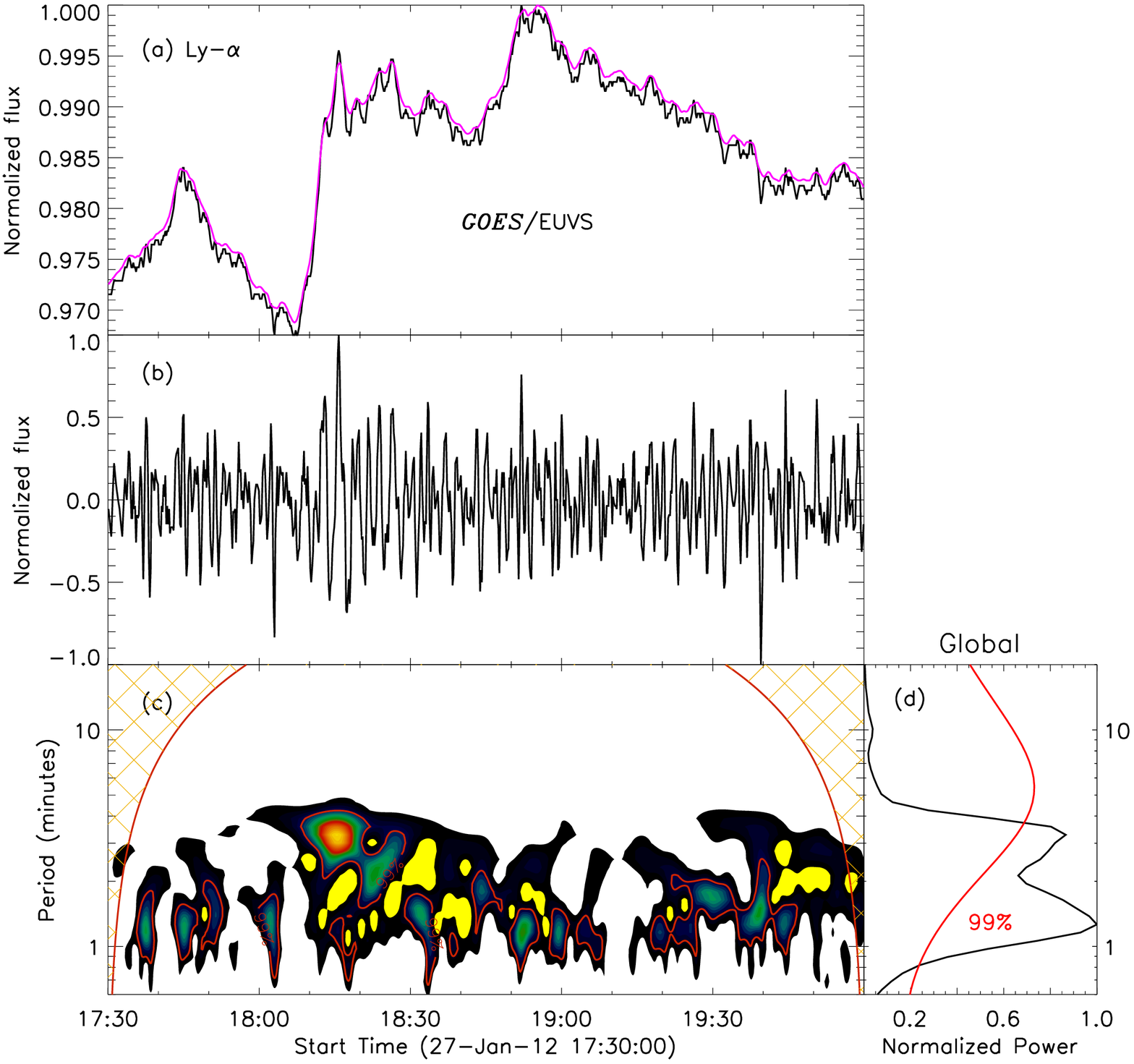}
\caption{Wavelet analysis result in Ly-$\alpha$ line of the X1.7
flare. Panel~(a): Normalized original (black) and trended (magenta)
light curves. Panel~(b): Normalized detrended light curve.
Panels~(c) \& (d): The wavelet power spectrum and the global wavelet
power. The red lines indicate significance levels of 99\%.
\label{lya1}}
\end{figure}

\begin{figure}
\centering
\includegraphics[width=\linewidth,clip=]{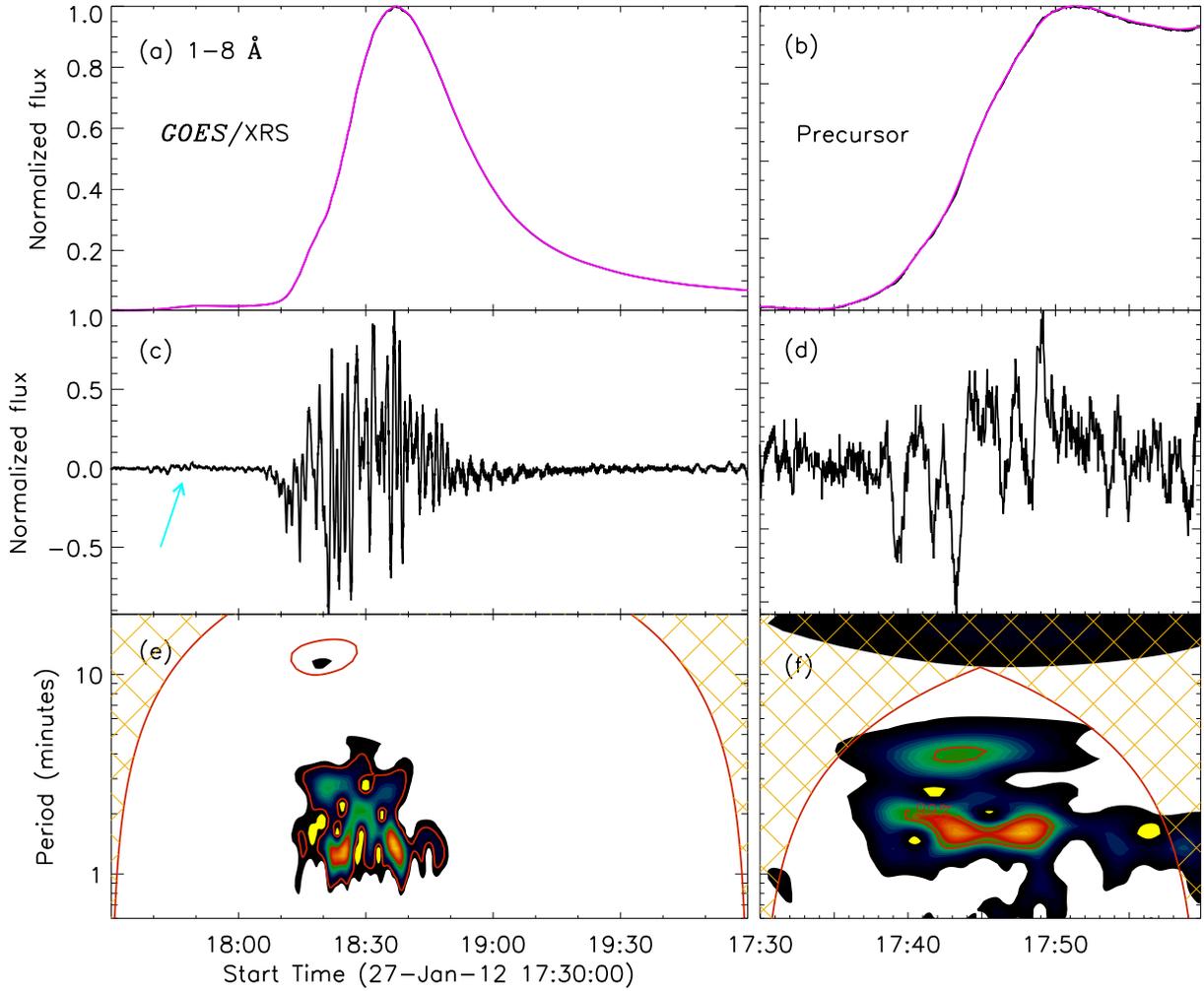}
\caption{Wavelet analysis result in {\it GOES}~1$-$8~{\AA} of the
X1.7 flare. Panels~(a)~\&~(b): Normalized original (black) and
trended (magenta) light curves. Panels~(c)~\&~(d): Normalized
detrended light curve. Panels~(e)~\&~(f): The wavelet power spectra.
The red lines indicate significance levels of 99\%. \label{goes1}}
\end{figure}

It is worthwhile to stress that the 1-minute QPP can be clearly
observed in the Ly-$\alpha$ emission from the precursor through
impulsive to gradual phases of the X1.7 flare, as shown in
Figure~\ref{lya1}. However, the QPP detected in the SXR flux during
the impulsive phase shows a much stronger signature when comparing
to that detected during the flare precursor phase, as seen in
Figure~\ref{goes1}. This could be associated with enhancements of
the flare radiation in different wavelengths. It can be seen that
the full-disk SXR irradiance during the flare impulsive phase is
much stronger than that during the flare precursor, i.e., nearly 100
times, as indicated by the red line in the panel~(e) of
Figure~\ref{flare1}. On the other hand, the Ly-$\alpha$ enhancement
during the solar flare is fairly weak. A statistical study of 477 M-
and X-class flares suggests that 95\% of these large flares show a
$\leq$10\% increase in the Ly-$\alpha$ emission above their
background, with a maximum enhancement of $\sim$30\%
\citep[e.g.,][]{Milligan19}. The similar Ly-$\alpha$ contrast during
solar flares is also discovered by \cite{Brekke96} and
\cite{Woods04}. Moreover, a much smaller increase such as $<$1\% in
Ly-$\alpha$ emission during solar flares is detected by the Large
Yield Radiometer \citep[see,][]{Kretzschmar13,Raulin13}. All these
findings are consistent with our observations, i.e., less than 10\%
enhancement compared to its background, as indicated with the black
line in Figure~\ref{flare1}~(e).

We address here the possible generation mechanism of the 1-minute
QPP in Ly-$\alpha$ emission. It could not depend on the nonthermal
electron produced by a periodic magnetic reconnection
\citep{Nakariakov09,Milligan17} due to the fact that it could be
detected from the flare precursor through impulsive to gradual phase
\citep[e.g.,][]{Tian16}. Conversely, given the fact that the
1-minute QPP can be discovered during the whole phases of solar
flares, it is most likely to be driven by a self-oscillation process
of the spontaneous magnetic reconnection, i.e., magnetic dripping
\citep[e.g.,][]{Nakariakov10,McLaughlin18}. On the other hand, it
might also be interpreted as the MHD wave
\citep{Reznikova11,Nakariakov16,Van16}. The 1-minute QPP in
full-disk SXR/HXR flux and local EUV emission has been found to
originate from the flare footpoints \citep{Ning17}, where the
Ly-$\alpha$ emission was supposed to be generated.
\citep{Allred05,Rubio09,Chamberlin18,Dominique18b}. Therefore, the
MHD oscillation at footpoints could also be used to explain the
1-minute QPP, i.e., the sausage wave
\citep[e.g.,][]{Tian16,Kolotkov18} or the acoustic wave
\citep[e.g.,][]{Milligan17,Milligan19}. However, it is impossible to
determine the mode of MHD wave, since the imaging observations in
Ly-$\alpha$ emission are very few. So, the oscillation location and
physical parameters such as the plasma density is still hard to be
determined. The future instruments such as the Extreme Ultraviolet
Imager \citep{Schuhle11} on board the Solar Orbiter
\citep{Marsch05}, the Lyman-$\alpha$ Solar Telescope
\citep{Feng19,Li19} aboard the Advanced Space-based Solar
Observatory \citep{Gan19,Huang19}, are promising to solve this
issue.

\begin{figure}
\centering
\includegraphics[width=\linewidth,clip=]{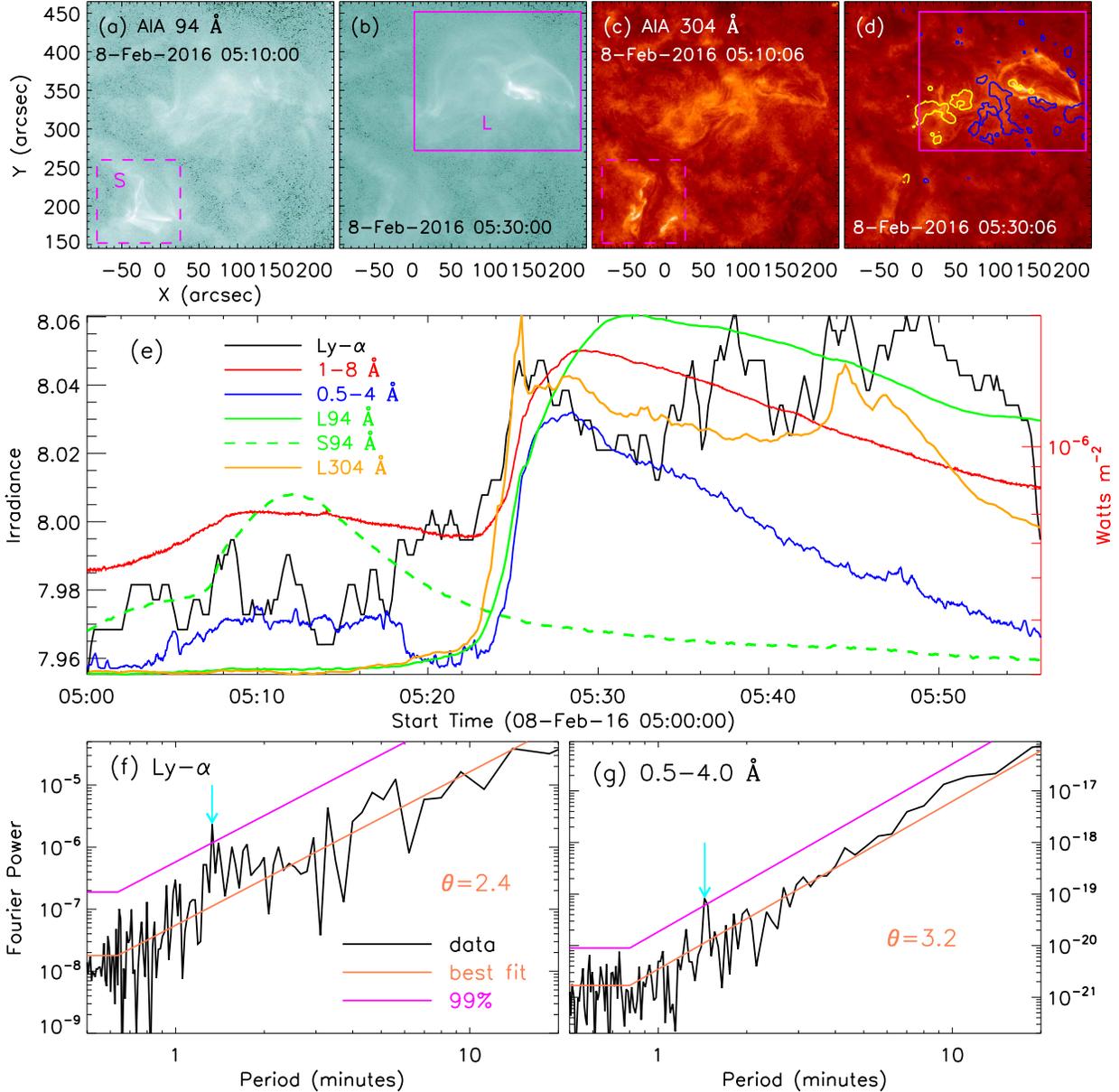}
\caption{Overview of solar flares on 2016 February 08.
Panels~(a)$-$(d): AIA images in 94~{\AA} and 304~{\AA}. The yellow
and blue contours represent the positive and negative magnetic
fields at the levels of $\pm$100~G. The magenta boxes mark the
bright regions used to integrate the local EUV flux. Panel~(e):
Full-disk flux in Ly-$\alpha$ (black), {\it GOES}~1$-$8~{\AA} (red),
and 0.5$-$4.0~{\AA} (blue), the local flux in AIA~94~{\AA} (green)
and 304~{\AA} (orange). Panels~(f)$-$(g): Fourier power spectra
derived from the Ly-$\alpha$ and {\it GOES}~0.5$-$4~{\AA}. The coral
and magenta lines indicate the best fit results and the confidence
levels at 99\%, respectively. The cyan arrow indicates the period
above the confidence level. \label{flare2}}
\end{figure}

\begin{figure}
\centering
\includegraphics[width=\linewidth,clip=]{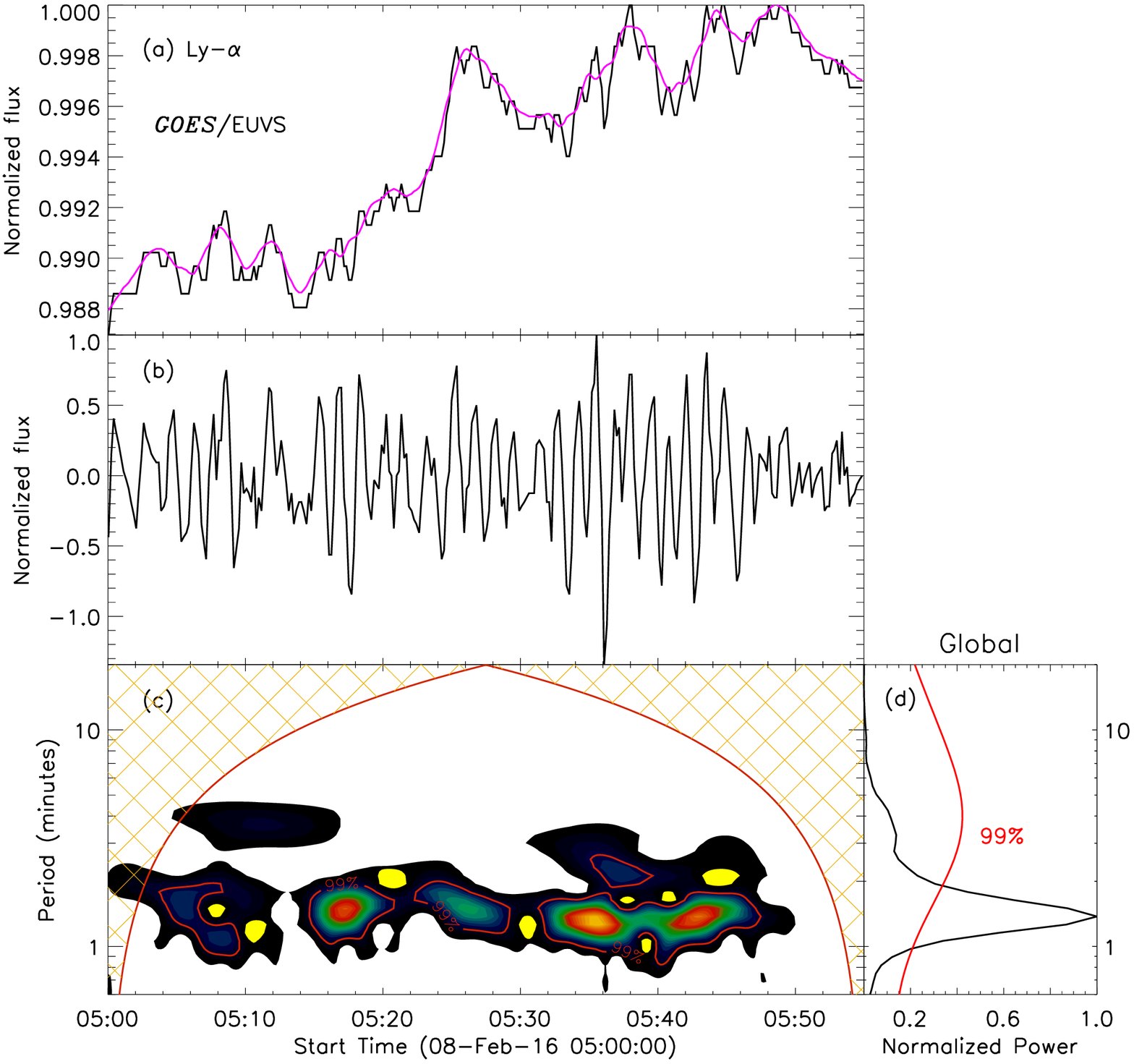}
\caption{Wavelet analysis result in Ly-$\alpha$ line of the C1.6
flare. Panel~(a): Normalized original (black) and trended (magenta)
light curves. Panel~(b): Normalized detrended light curve.
Panels~(c)~\&~(d): The wavelet power spectrum and the global wavelet
power. The red lines indicate significance levels of 99\%.
\label{lya2}}
\end{figure}

\begin{figure}
\centering
\includegraphics[width=\linewidth,clip=0]{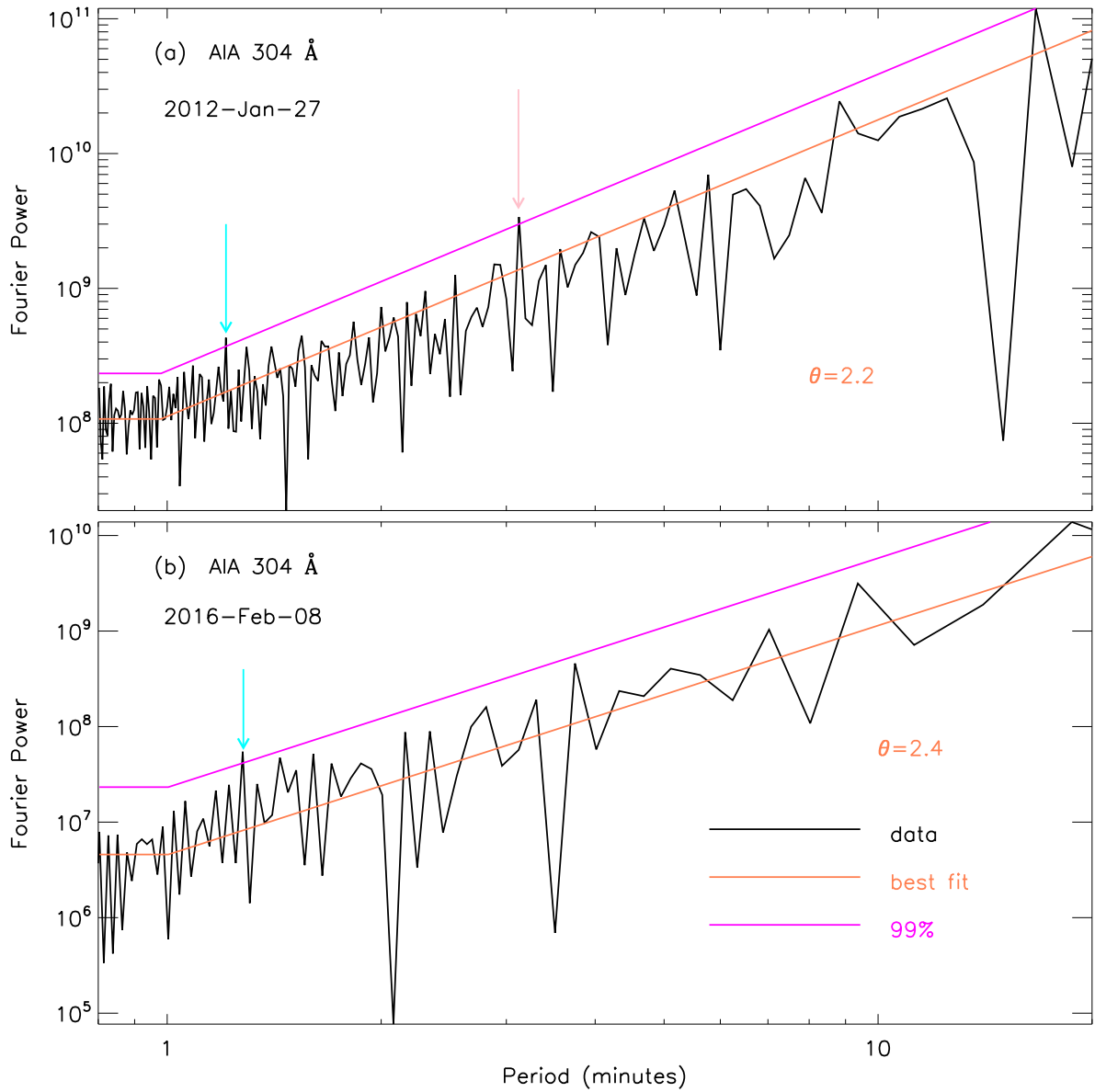}
\caption{Fourier spectra in AIA~304~{\AA} for the flares on 2016
January 27 (a) and 2016 February 08 (b), respectively. The coral and
magenta lines indicate the best fit results and the confidence
levels of 99\%. \label{aia}}
\end{figure}

Finally, we want to stress that the shorter period of the
flare-related QPP ranges between $\sim$1.1$-$1.4~minutes, while the
longer period varies from $\sim$3.2~minutes to $\sim$3.5~minutes, as
indicated by the peaks of the Fourier power spectra in
Figures~\ref{flare1}, \ref{flare2} and \ref{aia}. However, there are
diffuse period ranges in wavelet power spectra, as shown in
Figures~\ref{lya1}$-$\ref{lya2}. For simplicity, we regarded them as
1- or 3-minute periods in this paper, which were similar to previous
findings \citep[e.g.,][]{Milligan17,Ning17}. Using the {\it SDO}/AIA
observations, the QFP magnetosonic wave trains after a C2.0 flare
were found to have a common period of $\sim$80$\pm$10~s
\citep[see][]{Shen13}, which is similar to the 1-minute QPP we found
here.. Moreover, the quasi-periods both in their QFPs and our
flare-related QPPs were attributed to some periodic processes in
magnetic reconnection \citep[see also][]{Liu11,Yuan13,Kumar17}. As a
next-step work, we will search for QFP wave trains with a period of
$\sim$1~minute in the future.

\acknowledgments  We thank the anonymous referee for his/her
valuable comments. The authors would like to acknowledge Profs.
V.~Nakariakov, S.~Feng, H.~Tian, Y.~Ding, and J.~Chae, for their
inspiring discussions. We thank the teams of {\it GOES}/EUVS, {\it
GOES}/XRS, and {\it SDO}/AIA for their open data use policy. This
work is supported by NSFC under grants 11973092, 11921003, 11973012,
11873095, 11790302, 11790300, 11729301, 11773061, and U1731241,
U1931138, the Youth Fund of Jiangsu No. BK20171108, and KLSA202003,
as well as CAS Strategic Pioneer Program on Space Science, Grant No.
XDA15052200, XDA15320103, and XDA15320301.  D.~Li is also supported
by the Specialized Research Fund for State Key Laboratories. This
work is also supported by the mobility grant of Sino-German Science
Center M-0068, and National key research and development program
2018YFA0404202. The Laboratory No. 2010DP173032.

\begin{table}
\caption{The metrics of instruments used in this study.} \centering
\setlength{\tabcolsep}{20pt}
\begin{tabular}{c c c c c c c}
 \hline\hline
Instrument          &   Wavelength            &   Cadence (s)   &  Channel   \\
\hline
  {\it GOES}/EUVS   &   1216~{\AA}            &    $\sim$11     &   Ly-$\alpha$   \\
\hline
  {\it GOES}/XRS    &   1$-$8~{\AA}           &    $\sim$2      &   SXR         \\
                    &   0.5$-$4.0~{\AA}       &    $\sim$2      &   SXR         \\
\hline
 {\it SDO}/AIA      &     94~{\AA}            &     12          &   EUV        \\
                    &     304~{\AA}           &     12          &   EUV         \\
\hline\hline
\end{tabular}
\label{tabl}
\end{table}

\end{document}